\documentclass[a4paper,10pt,twoside]{cpc-hepnp}
\usepackage{CJK,upgreek,fancyhdr}
\usepackage{multicol}
\usepackage{graphicx}
\usepackage{booktabs}
\usepackage{amssymb,bm,mathrsfs,bbm,amscd}
\usepackage[tbtags]{amsmath}
\usepackage{lastpage}
\usepackage{hyperref}
\usepackage[tbtags]{amsmath}  %
\usepackage{amstext} %
\usepackage{balance} %
\usepackage{color,soul}
\usepackage{epstopdf}

\newcommand{\blue}[1]{{ \color{blue} #1 }}

\usepackage{soul}

\begin{document}


\fancyhead[C]{\small 
**} \fancyfoot[C]{\small **-\thepage}


\title{\boldmath
Shape Polarization and Quasiparticle Alignment in the $[523]5/2$ and $[642]5/2$ bands of $^{169}\text{Hf}$
\thanks{Supported by
National Natural Science
Foundation of China (No12205076,203070092510)}}

\author{%
     Rong-Xin Nie$^{1}$, Xue-Hui Ai$^{1}$,
     Xin Guan$^{1)}$\email{guanxin@lnnu.edu.cn},
     Jie  Yang$^{1}$}
\maketitle
\renewcommand{\headrulewidth}{0pt}

\address{%
$^1$ Department of Physics, Liaoning Normal University,
Dalian 116029, China\\
}

\begin{abstract}
   
The rotational properties of the $[523]5/2$ (originating from the $h_{11/2}$ shell) and $[642]5/2$ (originating from the $i_{13/2}$ shell) signature partner bands in the odd-$A$ nucleus $^{169}\text{Hf}$ are investigated using the Total Routhian Surace (TRS) method. Experimental observations reveal a distinct signature inversion in the $[523]5/2$ band at high spin, whereas the $[642]5/2$ band exhibits a larger, more conventional signature splitting. 
The study elucidates the microscopic origin of the signature inversion observed in the $[523]5/2$ ($h_{11/2}$) band at $I = 35/2$. Our analysis identifies a prominent proton subshell gap at $Z=72$ ($\beta_2 \approx 0.35$) that effectively "locks'' the proton core, allowing neutron-driven dynamics to dominate the signature staggering. A critical shape bifurcation is identified within the $[523]5/2$ configuration: while the $\alpha = -1/2$ branch remains rigid at high quadrupole deformation ($\beta_2 \approx 0.32, \gamma \approx -10^\circ$), the $\alpha = +1/2$ branch undergoes a transition toward a reduced $\beta_2 \approx 0.20$ and an enhanced hexadecapole deformation.
This structural shift facilitates a prompt $i_{13/2}$ neutron alignment at $\hbar\omega \approx 0.3$ MeV for the $\alpha = +1/2$ branch. Furthermore, the $\alpha = -1/2$ signature enters a highly $\gamma$-soft regime at high spin, with triaxiality fluctuating between $-10^\circ$ and $+10^\circ$. This excursion into positive $\gamma$ values energetically favors the unfavored signature in the rotating frame, reinforcing the Routhian crossing initiated by the $(\beta_2, \beta_4)$ polarization. 
In contrast, the $[642]5/2$ ($i_{13/2}$) band maintains a stable triaxial shape near $\gamma \approx -18^\circ$, preserving its signature splitting and preventing inversion. At the extreme frequency of $\hbar\omega \approx 0.5$ MeV, a high-spin "shape-jump" to $\beta_2 \approx 0.38$ is predicted for the $i_{13/2}$ band, signaling the breakthrough of the $Z=72$ shell gap and the onset of a highly deformed proton-aligned regime.
\end{abstract}




\begin{multicols}{2}

\section{Introduction}
In the framework of the collective model, signature is a fundamental quantum number used to classify high-spin rotational spectra. It is associated with the symmetry of the nuclear wave function under a $180^\circ$ rotation about an axis perpendicular to the symmetry axis \cite{Bohr1975}. 
For a deformed rotating nucleus, signature distinguishes between two sequences of states differing by one unit of angular momentum ($\Delta I = 1$) \cite{Voigt1983}. These sequences, termed signature partner bands ($\alpha = \pm 1/2$ for odd-$A$ nuclei), arise from the intrinsic symmetry properties of single-particle orbitals under rotation. The energy separation between these partners, signature splitting, serves as a sensitive probe of nuclear shape, quasiparticle configurations, and rotational frequency \cite{Granderath1996}.
The magnitude of signature splitting is fundamentally linked to the admixture of $K=1/2$ components in the quasiparticle wave functions \cite{Mueller1994,Hamamoto19901,FRXu2000}. Consequently, large splitting is typically observed in low-$K$ orbitals within high-$j$ shells, whereas mid- to high-$K$ orbitals exhibit significantly reduced splitting \cite{Stephens1975,Myer-ter-Vehn1975}. Furthermore, this splitting is highly sensitive to triaxiality, as the chemical potential's position relative to the shell determines the specific $\gamma$ deformation induced by the valence quasiparticles \cite{Aryaeinejad1984}. 
A particularly intriguing phenomenon in this regime is signature inversion, where the expected energy ordering of the signature partners reverses. While frequently observed in odd-odd nuclei—such as the $\pi h_{11/2} \otimes \nu i_{13/2}$ configurations near $A \approx 160$ \cite{FRXu2000}—it also appears in odd-mass isotopes, challenging the predictions of the standard Cranked Shell Model (CSM). Theoretical interpretations of this crossover often invoke beyond-mean-field effects, including triaxial vibrations \cite{Ikeda1990}, wobbling motion \cite{Matsuzaki1992}, and quadrupole pairing correlations \cite{FRXu2000}.
The neutron-deficient hafnium isotopes ($Z=72$) provide a fertile testing ground for these effects, as they occupy a transitional region where the nuclear potential is remarkably soft~\cite{Schmidt2001}. 
In $^{169}\text{Hf}$, the $[523]5/2$ ($h_{11/2}$) band exhibits a pronounced signature inversion at $I = 35/2$, whereas the neighboring $[642]5/2$ ($i_{13/2}$) band maintains a large, conventional splitting. While such inversions in the $A \approx 170$ region are traditionally attributed to triaxiality ($\gamma$) or quadrupole ($\beta_2$) polarization~\cite{Bengtsson1986}, 
the abrupt nature of the crossing in $^{169}\text{Hf}$ suggests a more complex microscopic driver. In the present work, we employ Total Routhian Surface (TRS) calculations to demonstrate that this inversion is governed by a shape bifurcation. We aim to show that the interaction between a stable proton core, a signature-dependent hexadecapole ($\beta_4$) "stretch," and a transition into a $\gamma$-soft regime provides the necessary mechanism to reproduce the experimental alignment and signature splitting patterns.

\section{The theoretical framework}
The total Routhian surface (TRS) calculation applied in this work is based on the pairing deformation self-consistent cranked shell model (CSM)~\cite{FRXu2000,Satula1994}. The total nuclear energy $E^\omega$ comprises a macroscopic component and the microscopic correction. 
\begin{equation}
   E^\omega = E_{\text {mac}}(\omega, \beta) + \delta E_{\text {shell}}(\omega, \beta) + \delta E_{\text {pair}}(\omega, \beta).
                                                                                                                           \label{Eqn.0}
\end{equation}
The macroscopic energy $E_{\text{mac}}(\omega, \beta)$ is derived from the shape-dependent liquid-drop model (LDM), with parameters adopted from Myers and
Swiatecki~\cite{Myers1966}. The microscopic correction, which reflects the non-uniform distribution of single-particle levels in the nucleus, primarily includes shell and pairing corrections. These are computed via the Strutinsky smoothing method~\cite{Strutinsky1967} and the Lipkin-Nogami (LN) method~\cite{Lipkin1960,Nogami1964}, respectively.  

In the cranking model, the nucleus rotates around a fixed axis (usually the $x$-axis) at a given rotational frequency. Pairing correlations, which depend on frequency and deformation, are treated within the Hartree-Fock-Bogoliubov cranking (HFBC) approximation. The HFBC equations are solved self-consistently at each rotational frequency and at every point on the deformation grid, which includes quadrupole ($\beta_2$), triaxial ($\gamma$), and hexadecapole ($\beta_4$) shape degrees of freedom, ensuring pairing self-consistency throughout the calculations.
 
To improve the pairing description, the Lipkin-Nogami method is incorporated, resulting in cranked Lipkin-Nogami (CLN) equations that describe the nucleus in terms of independent quasiparticle energies, known as Routhians~\cite{Voigt1983,Satula1994,Pradhan1973}. The auxiliary Routhian $\hat{\mathcal{H}}$ is defined by 
\begin{equation}
   \hat{\mathcal{H}} = \hat{H}^{\omega}-\lambda_{1}\hat{N} - \lambda_{2}\hat{N}^2,
                                                                                                                           \label{Eqn.1}
\end{equation}
where $\hat{N}$ is the nucleon number operator and $\hat{H}^{\omega}$ is the cranking Hamiltonian, which contains a single-particle Hamiltonian, $\hat{H}_{\rm sp}$, a pairing Hamiltonian $ \hat{H}_{\rm pair}$, { and the so called cranking term depending on rotational frequency}, 
\begin{eqnarray}
     \hat{H}^\omega&=&\hat{H}_{\rm sp} + \hat{H}_{\rm pair} -\omega\hat{\jmath}_{x}  \nonumber  \\[2mm]
               &=&\sum_{k}e_{k}a^{+}_{k}a_{k} 
                  - G\sum_{k,l>0}a^{+}_{k}a^{+}_{\bar{k}} a_{\bar{l}}a_{l} -\omega\hat{\jmath}_{x},
                                                                                                                           \label{Eqn.2}
\end{eqnarray}
in which $e_{k}$ is the single-particle energy, $G$ is the pairing strength, and 
$|\bar{k}\rangle = \hat{T} | k \rangle$ and $\hat{T}$ is the time reverser operator, $\hat{\jmath}_{x}$ denotes the component of the total nucleonic angular momentum  operator, and $\omega$ the rotation frequency. The particle creation and annihilation operators are defined as $a^{+}$ and $a$.

In the Lipkin-Nogami method, the expectation value of $\hat{\mathcal{H}} $ is minimized assuming that coefficients $\lambda_{1}$ and $\lambda_{2}$ are constant: 
\begin{eqnarray}                                                                                                                          
     \delta\langle \Phi | \hat{\mathcal{H}}  | \Phi \rangle =0,
                                                                                                                           \label{Eqn.3}
\end{eqnarray}
in which $|\Phi\rangle$ takes the standard form of the BCS wave function as follows
\begin{eqnarray}                                                                                                                          
     |\Phi\rangle
     =
     \prod_{k}
     (u_{k}                    
       +
       v_{k} a^{+}_{k} a^{+}_{\bar{k}})
       | 0 \rangle ,
                                                                                                                           \label{Eqn.4}
\end{eqnarray}
with the normalisation condition 
\begin{eqnarray}                                                                                                                          
     u^{2}_{k}+ v^{2}_{k} = 1,
                                                                                                                           \label{Eqn.5}
\end{eqnarray}
in which $| 0 \rangle$ denotes the particle vacuum state, $v^2_{k}$ and $u^2_{k}$ are the particle occupied and unoccupied probability, respectively. 

The Lagrange multipliers $\lambda$($\lambda_1$, $\lambda_2$) and $\omega$ are determined by fixing expectation values of the particle number and angular momentum, respectively. 
\begin{eqnarray}                                                                                                                          
    \langle \Phi | \hat{N} |\Phi\rangle =n, \quad n=Z\; \text {or } N.
                                                                                                                           \label{Eqn.6}
\end{eqnarray}
and
\begin{eqnarray}                                                                                                                          
    \langle \Phi | \hat{j}_{x} |\Phi \rangle =I_{x}.
                                                                                                                           \label{Eqn.7}
\end{eqnarray}
The total angular momentum along the rotation axis, $I_x$, is determined self-consistently within the cranking framework. Specifically, at each deformation $(\beta_2, \gamma,\beta_4)$ on the Total Routhian Surface, $I_x$ is extracted as the expectation value of the cranking operator, $\langle \hat{J}_x \rangle$. 

The quantity $\lambda_{1}$ represents the Lagrange multiplier used to constrain the average particle number, the parameter $\lambda_{2}$ is determined from the following auxiliary equation for the particle number fluctuations,
\begin{eqnarray}    
     \lambda_{2} = 
     \frac{G}{4}
     \frac{\sum_{k>0}(u^3_{k}v_{k})\sum_{k>0}(u_{k}v^3_{k}) - \sum_{k>0}(u_{k}v_{k})^4}
            {\sum_{k>0}(u^2_{k}v^2_{k})^2 - \sum_{k>0}(u_{k}v_{k})^4}.
                                                                                                                               \label{Eqn.8}                                                                                                                               
\end{eqnarray}
The single-particle energies required for calculating the shell correction are derived from the deformed Woods-Saxon (WS) potential with the universal parametrization~\cite{Nazarewicz1987}. In the diagonalization of the WS Hamiltonian, oscillator basis states with principal quantum numbers up to $N \leq 12$ for protons and $N \leq 14$ for neutrons are included. 

The nuclear shape is specified using the standard parametrization expanded in spherical harmonics up to second order~\cite{Nazarewicz1985}. Following the Lund convention~\cite{Andersson1976}, Cartesian quadrupole coordinates are employed in the calculations to parameterize the quadrupole deformation, explicitly incorporating the $\gamma$ degree of freedom as
\begin{equation}
  X=\beta_2\cos(\gamma+30^{\circ})
 \quad{\rm and}\quad 
 Y=\beta_2\sin(\gamma+30^{\circ})
                                                                                                                               \label{Eqn.9}                                                                                                                               
\end{equation}
The $\gamma$ parameter covers the range $-120^{\circ}$ to $60^{\circ}$.

In reflection-symmetric nuclei, both signature ($r$) and intrinsic parity ($\pi$) are good quantum numbers.
By definition $r$ is the eigenvalues of signature operator $\hat{\mathcal{R}}=e^{-i\pi J_x}$ and introduce the notation~\cite{Bohr1975,Voigt1983}
\begin{equation}
  r \stackrel{\mathrm{def.}}{=} e^{ (-i \pi \alpha)},
                                                                                                                               \label{Eqn.10}                                                                                                                               
\end{equation}
 and one finds  
\begin{equation}
  r =(-1)^{I},
                                                                                                                               \label{Eqn.11}                                                                                                                               
\end{equation}
where $I$ denotes the total nuclear spin. Thus for an even number of nucleons 
\begin{eqnarray}                                                                                                                          
   &&r=+1  (\alpha = 0),
   \quad\quad 
   I=0,2,4,... \\[1mm]
                                                                                                                               \label{Eqn.12}
   &&r=-1  (\alpha = 1),
   \quad\quad
   I=1,3,5,...
                                                                                                                               \label{Eqn.13}                                                                                                                           
\end{eqnarray}
For an odd number of nucleons 
\begin{eqnarray}                                                                                                                          
  && r=-i  (\alpha = \textstyle + \frac{1}{2}),
   \quad\quad \textstyle
   I=\frac{1}{2},\frac{5}{2},\frac{9}{2}, ... \\[1mm]
                                                                                                                               \label{Eqn.14}
   &&r=+i   (\alpha = \textstyle - \frac{1}{2}),
   \quad \quad  \textstyle
   I=\frac{3}{2},\frac{7}{2},\frac{11}{2}, ...
                                                                                                                               \label{Eqn.15}                                                                                                                           
\end{eqnarray}
Consequently, the invariance of the intrinsic Hamiltonian under signature symmetry results in two rotational bands that differ by $1 \hbar$ in spin.  Each rotational band is further split by signature symmetry into two families of levels differing by $2 \hbar$ in spin. These signature partner bands share the same intrinsic configuration but differ in their signature eigenvalue $r$. Commonly, the signature exponent $\alpha$—defined by $I \equiv \alpha \pmod{2}$—is used to label these bands. When the odd particle occupies a high-$j$ orbital, the favored signature, defined as
\begin{equation}
  \alpha_{f} = \frac{1}{2} (-1)^{j-\textstyle\frac{1}{2}},
                                                                                                                               \label{Eqn.16}                                                                                                                               
\end{equation}
and the unfavored signature is 
\begin{equation}
  \alpha_{u} = \frac{1}{2} (-1)^{j+\textstyle\frac{1}{2}},
                                                                                                                               \label{Eqn.17}                                                                                                                               
\end{equation}
where $j$ denotes the total angular momentum of the odd particle~\cite{hamamoto1990}.

\section{Results and discussions}
\label{Sect.01}
The variation in odd-even staggering observed in these bands supports the interpretation that the effect is related to the signature dependence of the Coriolis interaction. The staggering can be characterized by the energy staggering parameter $S(I)$, which is defined as follows~\cite{Bengtsson1984}:
\begin{eqnarray}
  S(I)=E(I)-[E(I+1)+E(I-1)]/2.
                                                                                                                               \label{Eqn.18}
\end{eqnarray}
\begin{center}
\includegraphics[width=7.5cm]{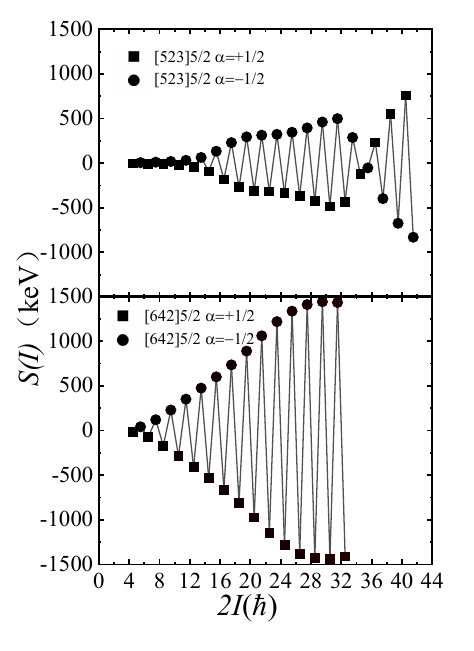}
\figcaption{Energy staggering parameter $S(I)$ as a function of spin $2I$ for the $[523]5/2$ ($h_{11/2}$) and $[642]5/2$ ($i_{13/2}$) bands in $^{169}_{\;\;\;72}\text{Hf}_{97}$. Squares and circles represent the $\alpha = +1/2$ and $\alpha = -1/2$ signature branches, respectively. The $[642]5/2$ band exhibits significantly larger splitting, whereas the $[523]5/2$ band displays a signature inversion at $2I = 35$ following the neutron alignment. Experimental data sources are taken from~\cite{Schmidt2001}.
                                                                                                                           \label{Fig_SI}
}
\end{center}

The energy staggering parameters $S(I)$ for the $[523]5/2$ ($h_{11/2}$) and $[642]5/2$ ($i_{13/2}$) bands in $^{169}_{\;\;\;72}\text{Hf}_{97}$. are presented as a function of spin $2I$ in Fig.~\ref{Fig_SI}. A striking contrast in the signature evolution of these two configurations is observed. 
The $[642]5/2$ band displays a large, persistent signature splitting that does not invert, even at high rotational frequencies. This behavior is characteristic of high-$j$, low-$\Omega$ orbitals, where a strong Coriolis coupling between the odd nucleon and the core rotation leads to a significant energy separation between the $\alpha = +1/2$ and $\alpha = -1/2$ branches~\cite{Granderath1996}. 
In contrast, the $[523]5/2$ band exhibits a much smaller staggering at low spins, followed by a distinct signature inversion around $2I = 35$. 

\begin{center}
\includegraphics[width=8.8cm]{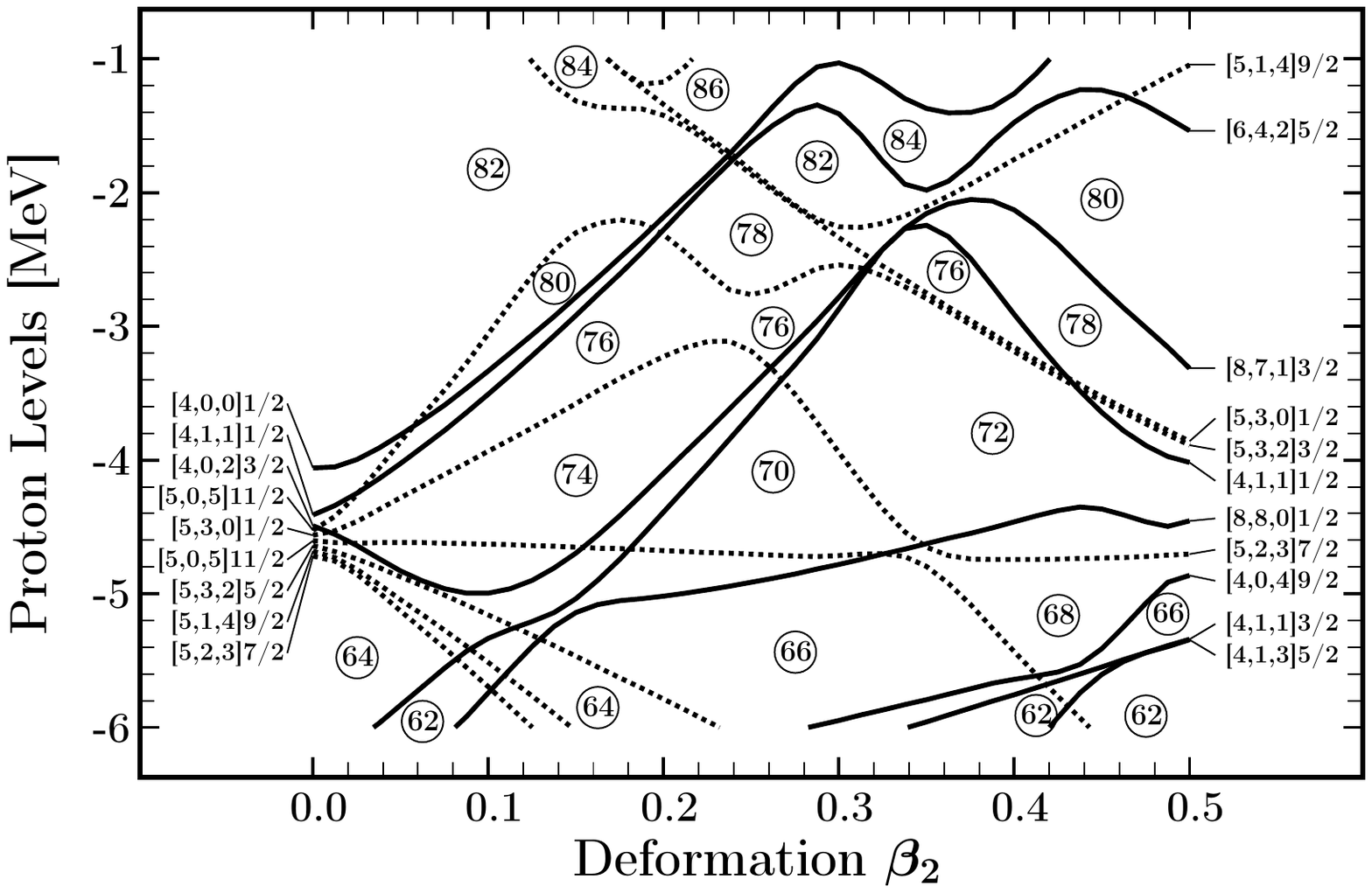}
\includegraphics[width=8.8cm]{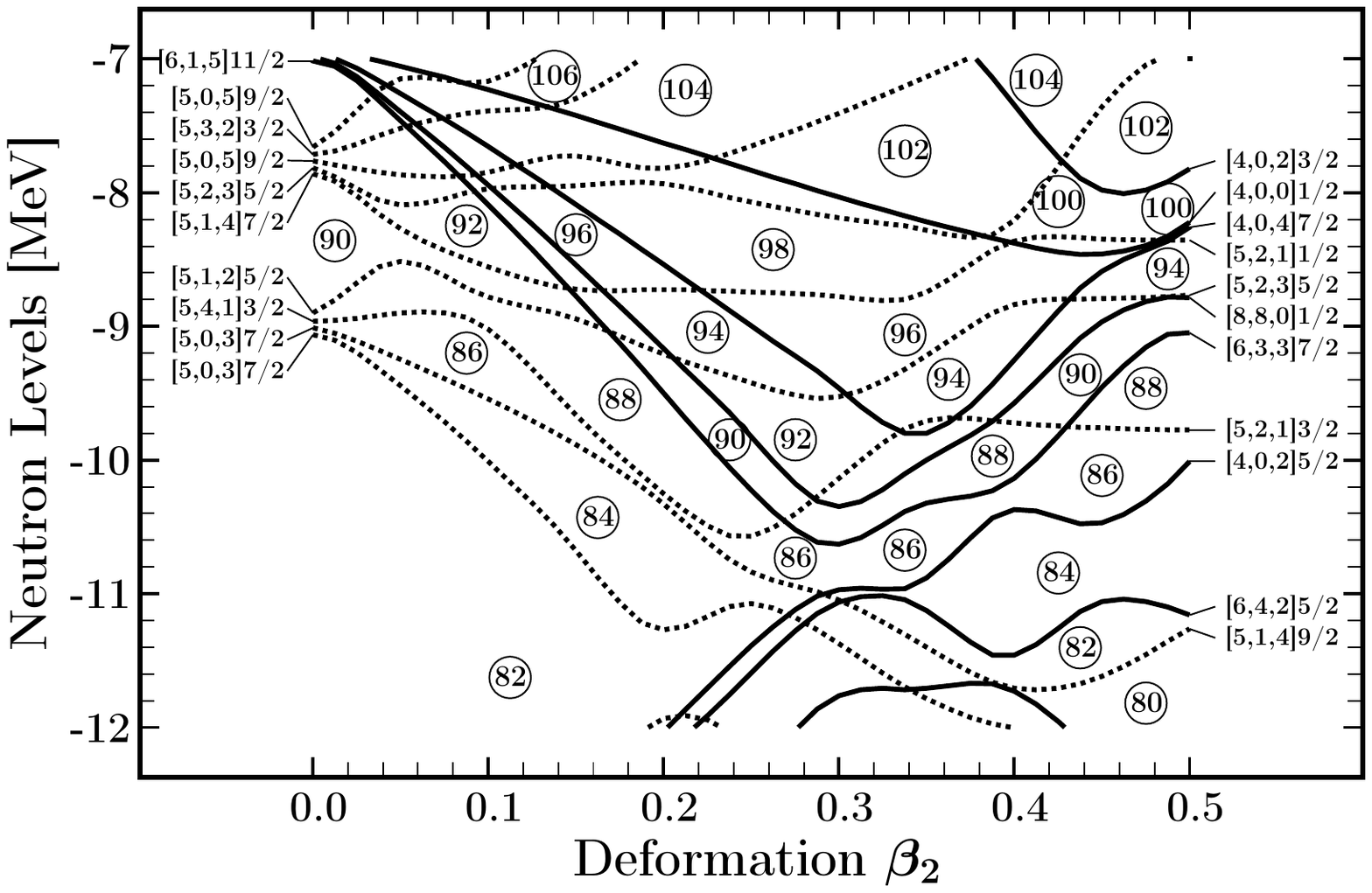}
\figcaption{Proton and Neutron single-particle energy spectra as a function of the quadrupole deformation parameter $\beta_2$.  A pronounced shell gap is observed at $Z=72$ for $\beta_2 \approx 0.35$ (top) and the distinct gaps $N=96$ appear at $\beta_2 \approx 0.20$ and $\beta_2 \approx 0.30$ (bottom).  These gaps correspond to the stabilized deformations for the $[642]5/2$ and $[523]5/2$ rotational configurations.
                                                                                                                           \label{Fig_SPE}
}
\end{center}

To study the microscopic origin of this anomalous staggering and the high-spin inversion, we examine the evolution of the single-particle energy spectra as a function of the quadrupole deformation parameter $\beta_2$, as shown in Fig.~\ref{Fig_SPE}. For the proton system, a prominent shell gap is observed at $Z=72$ for $\beta_2 \approx 0.35$,  while for the neutron system ($N=97$), distinct shell gaps appear at $\beta_2 \approx 0.2$ and $\beta_2 \approx 0.3$. The $[642]5/2$ orbital is found to be the primary occupation near the $\beta_2=0.2$ gap, consistent with the smaller deformation and larger signature splitting observed experimentally for this band. In contrast, at the larger deformation of $\beta_2 = 0.3$, the $[523]5/2$ orbital descends to the Fermi surface, becoming the energetically favored configuration. 

To validate the proposed shape evolution and alignment mechanisms, the experimental aligned angular momentum $I_x$ is compared with the TRS predictions in Fig.~\ref{Fig_Ixet}. Overall, the theoretical calculations show good agreement with the experimental data, successfully reproducing the alignment frequencies and the magnitude of the aligned gain for $[523]5/2$ band. For the $[642]5/2$ band, the TRS calculation reproduces the overall trend of the experimental alignment, although a small, sharp increase in $I_x$ is predicted at $\hbar\omega \approx 0.25$ MeV, which agrees within \cite{Miller2019,Yu2008}. 

\begin{center}
\includegraphics[width=9.6cm]{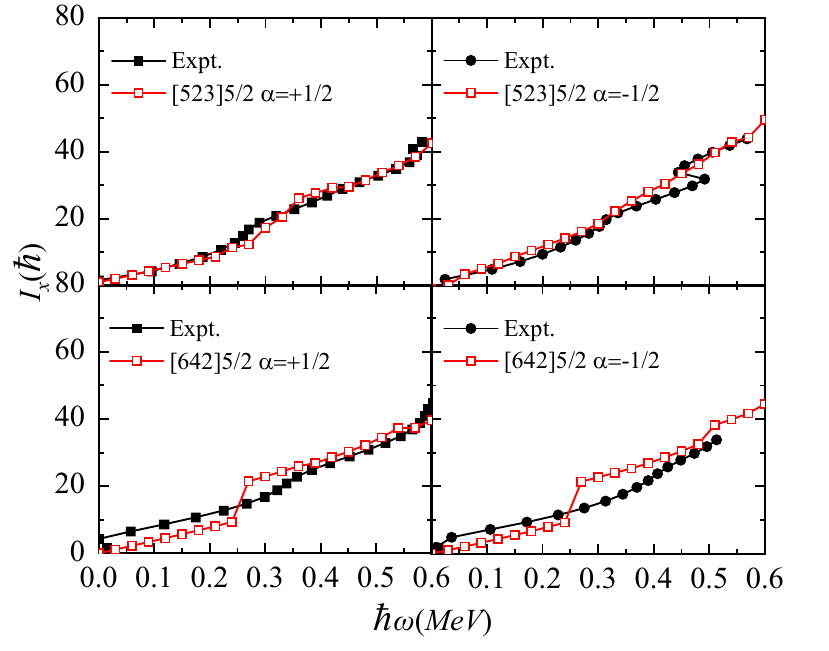}
\figcaption{
Aligned angular momentum $I_x$ as a function of rotational frequency $\hbar\omega$ for the $[523]5/2$ and $[642]5/2$ bands in $^{169}_{\;\;\;72}\text{Hf}_{97}$. Experimental data (black dots) are compared with the results of the self-consistent Total Routhian Surface (TRS) calculations (red squares for $\alpha=+1/2$ and blue circles for $\alpha=-1/2$ signature bands). The theoretical results provide a good reproduction of the measured alignments.
                                                                                                                           \label{Fig_Ixet}
}
\end{center}

The TRS maps at $\hbar\omega = 0.3$ MeV (Fig.~\ref{Fig_TRS}) provide a visual landscape of the competing nuclear shapes. A particularly intriguing feature is observed in the $[523]5/2$ ($\alpha = +1/2$) configuration. Although the equilibrium deformation is found at $\beta_2 \approx 0.25$ with axial symmetry ($\gamma = 0^\circ$), there exists a well-defined secondary minimum at $\gamma \approx -15^\circ$. The energy barrier separating these two minima is remarkably low, suggesting a high degree of $\gamma$-softness. The calculated quadrupole deformations for both the $[523]5/2$ configurations are quantitative agreement with the values reported by Dracoulis et al. \cite{Dracoulis2016}.
This shape coexistence suggests that the $[523]5/2$ band is not a rigid rotor but rather a highly polarizable system. The existence of the $\gamma \approx -15^\circ$ minimum, which is the dominant shape in the $[642]5/2$ band, indicates that the $h_{11/2}$ neutron orbital in this signature branch is struggling to maintain its axial character against the triaxial-driving tendencies of the $i_{13/2}$ intruder orbitals.

\end{multicols}
\begin{center}
\includegraphics[width=8.8cm]{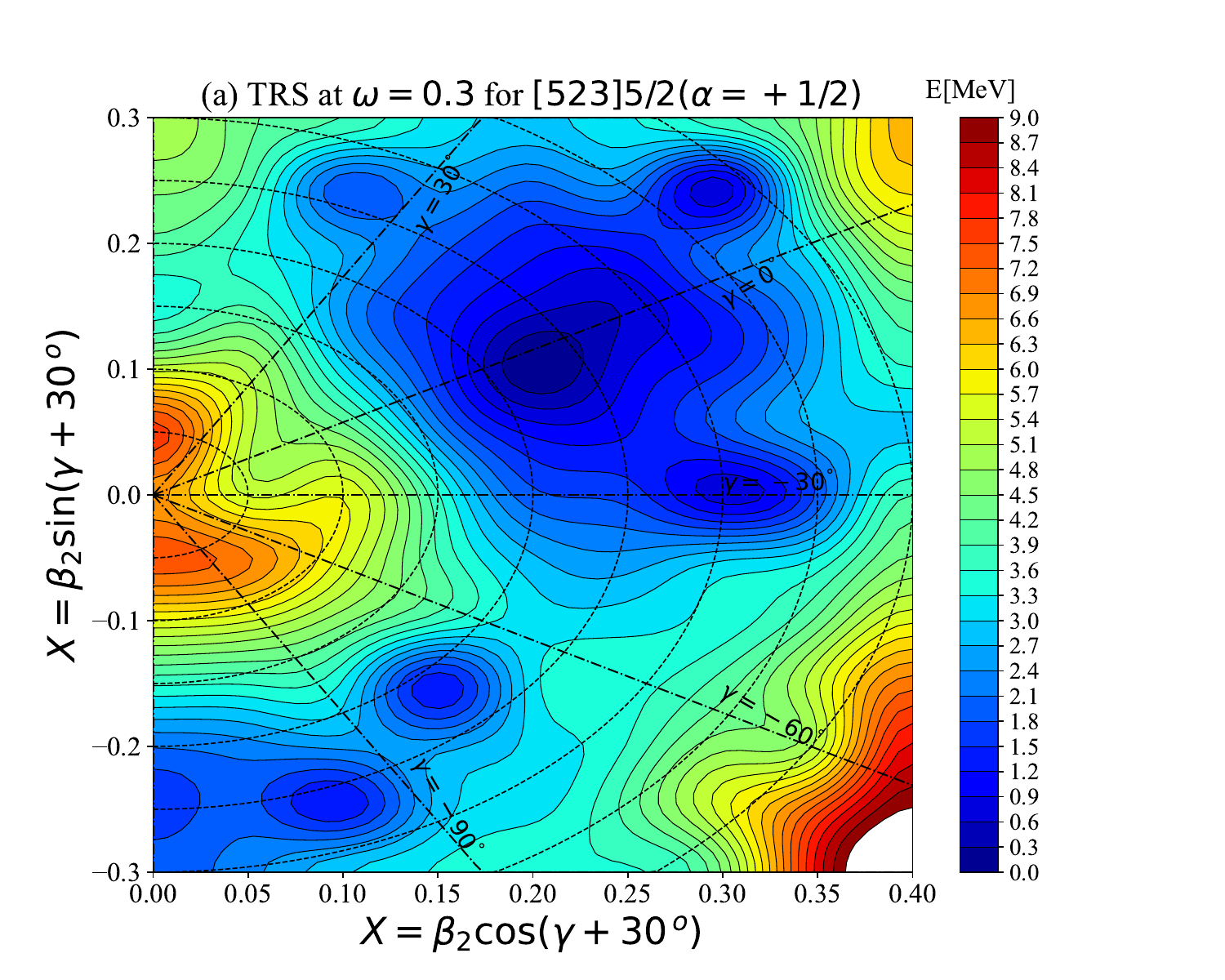}
\includegraphics[width=8.8cm]{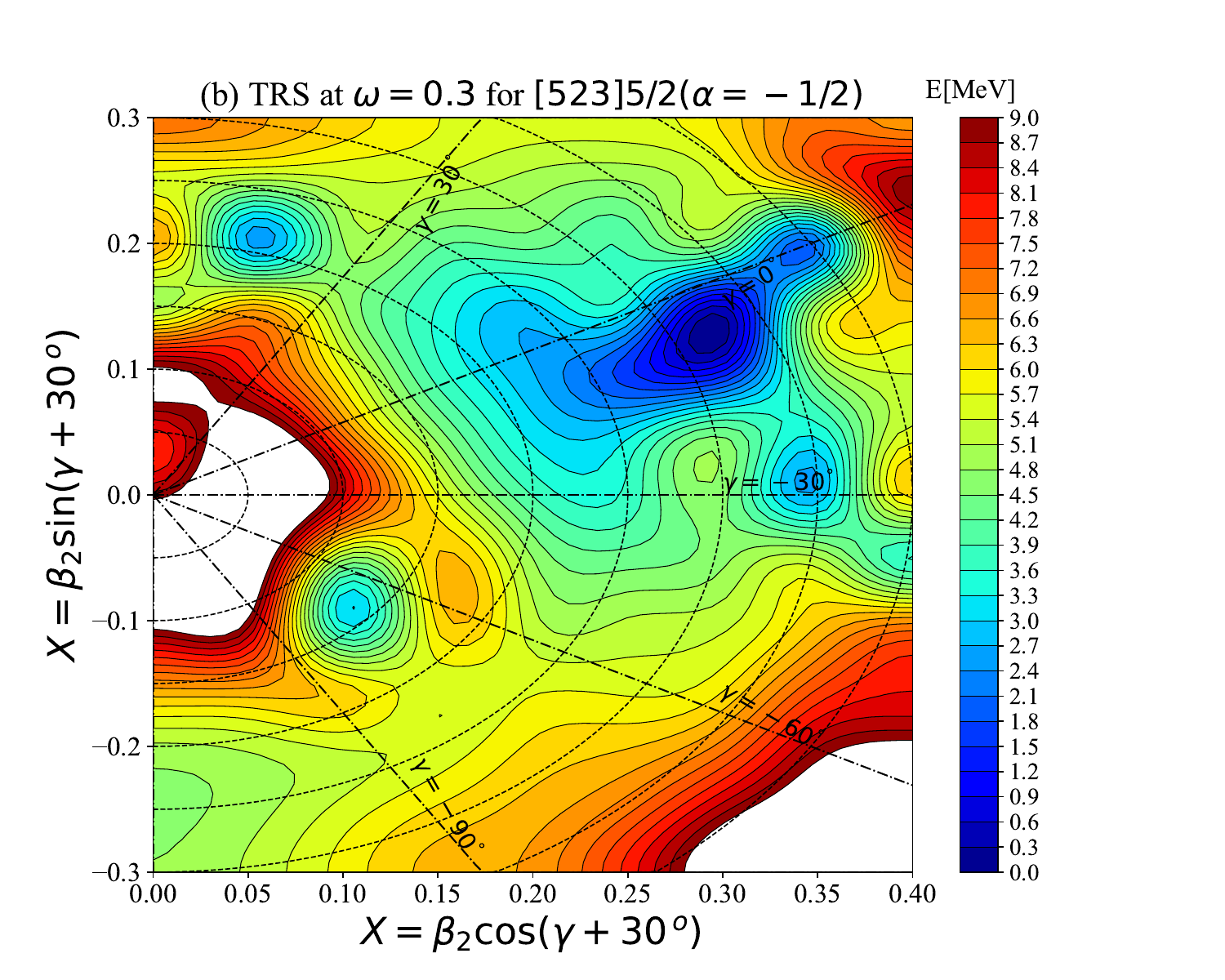}
\includegraphics[width=8.8cm]{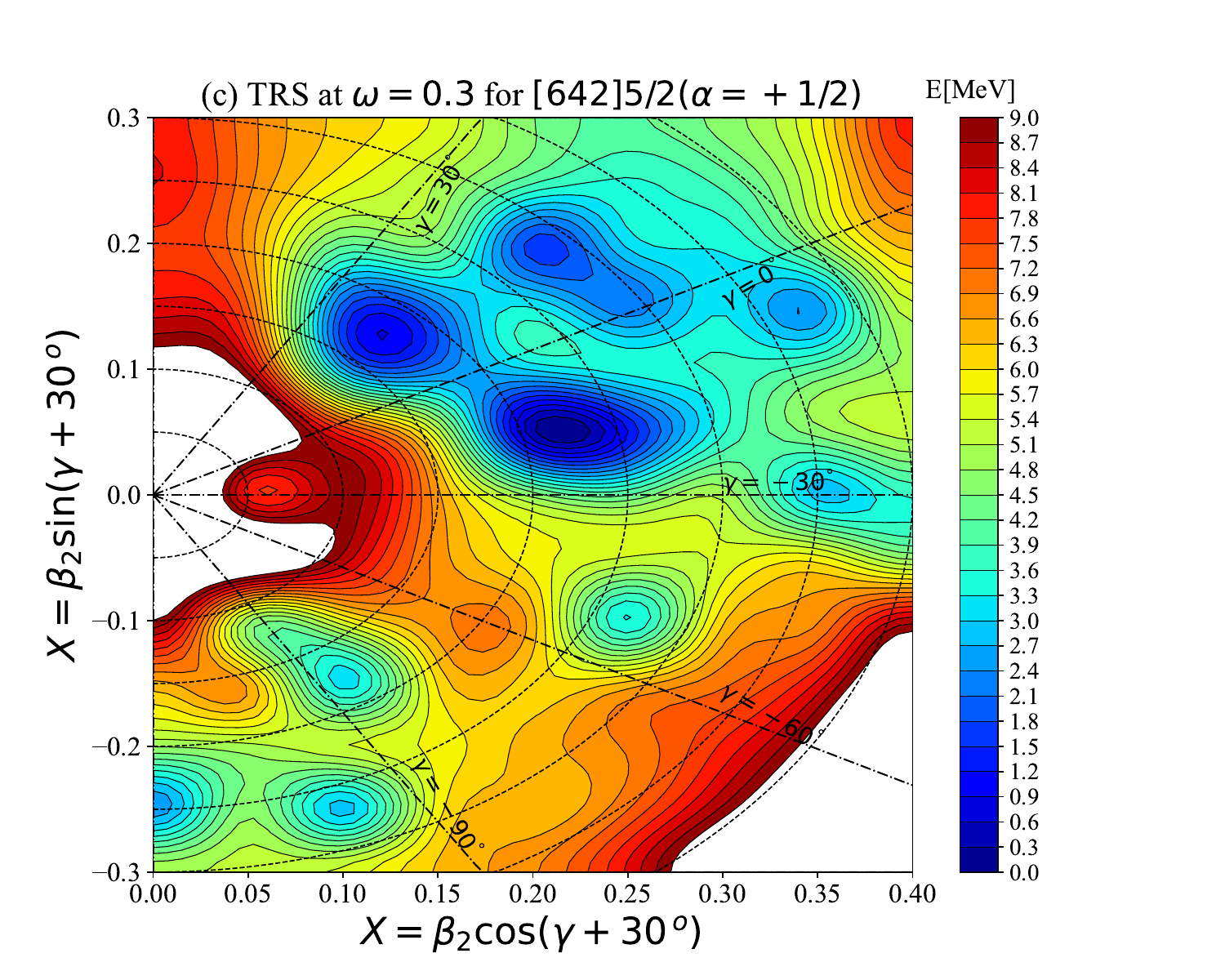}
\includegraphics[width=8.8cm]{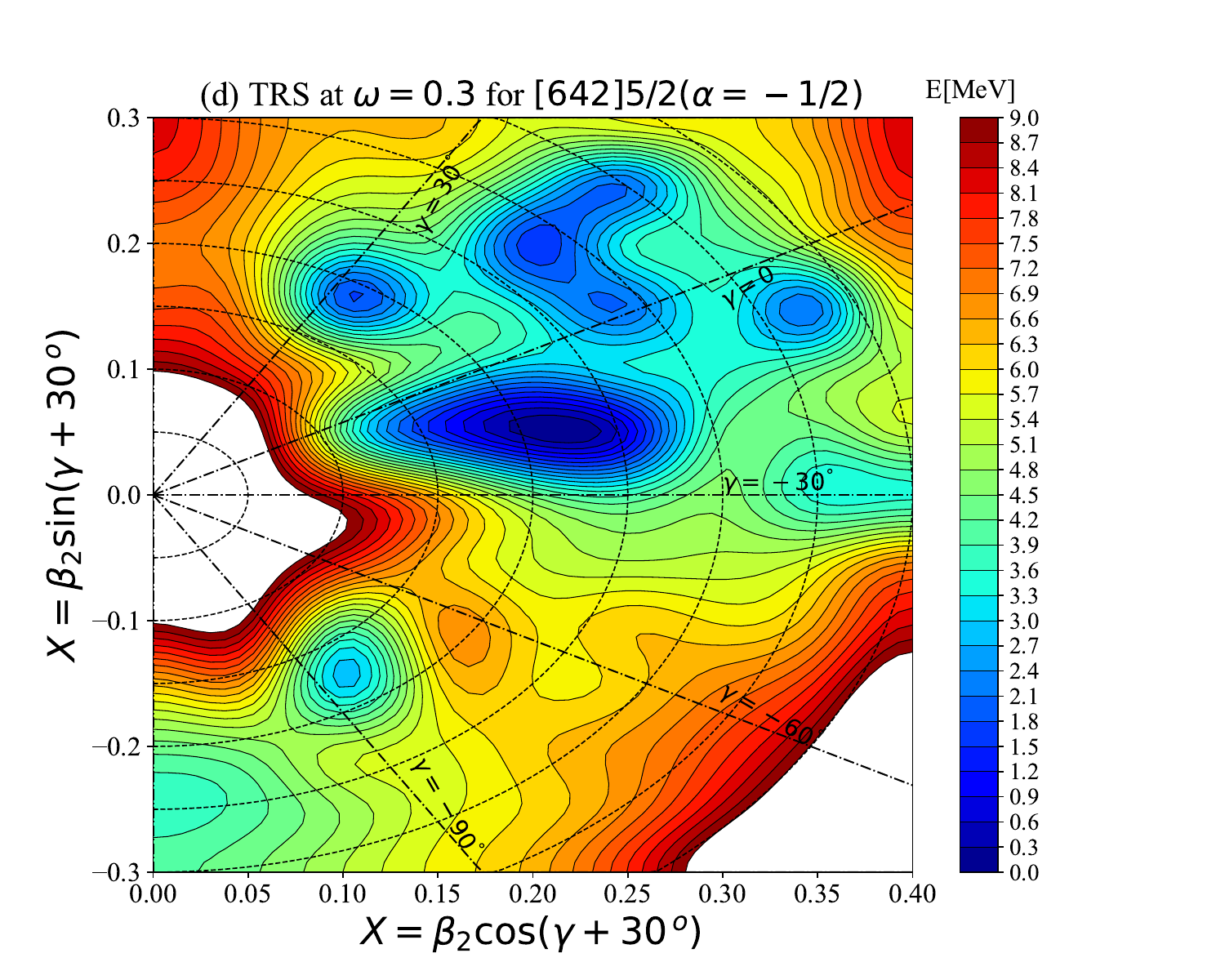}
\figcaption{Total Routhian Surfaces (TRS) for the  (a)  $\alpha = +1/2$ and  (b) $\alpha = -1/2$ signatures of the $[523]5/2$ band, and the (c) $\alpha = +1/2$ and (d) $\alpha = -1/2$ signatures of the $[642]5/2$ band in $^{169}_{\;\;\;72}\text{Hf}_{97}$ at a rotational frequency of $\hbar\omega = 0.3$ MeV. Contours are spaced at 300 keV intervals.                                                                                                                            \label{Fig_TRS}
}
\end{center}

\begin{multicols}{2}
 The microscopic origin of the alignment gain is elucidated by decomposing the total aligned angular momentum into its constituent proton and neutron components, as illustrated in Fig.~\ref{Fig_IpIn}. This decomposition reveals a contrast in the rotational response of the two configurations. 
 In the $[523]5/2$ ($h_{11/2}$) band, the proton alignment remains remarkably stable across a broad frequency range. This rigidity confirms that the $Z=72$ subshell gap at high quadrupole deformation ($\beta_2 \approx 0.35$), identified in the single-particle spectra (Fig.~\ref{Fig_SPE}), effectively stabilizes the proton core against Coriolis-induced pair breaking. 
 Consequently, the alignment dynamics are dominated by the neutron sector. A sharp increase in $I_n$ is observed for the $\alpha = +1/2$ branch at $\hbar\omega \approx 0.3$ MeV, whereas the $\alpha = -1/2$ partner exhibits a much more gradual alignment trajectory. This signature-dependent neutron alignment rate is the primary driver of the signature inversion observed in this band.
 
Conversely, the $[642]5/2$ ($i_{13/2}$) configuration follows a more symmetric evolution. The TRS calculations predict a primary $i_{13/2}$ neutron alignment at $\hbar\omega \approx 0.25$ MeV that occurs nearly simultaneously for both the $\alpha = +1/2$ and $\alpha = -1/2$ branches. 
Because both signature partners gain alignment at the same frequency, their respective Routhians ($e'$) decrease at comparable rates, thereby preserving the signature splitting and preventing a crossover. At higher rotational frequencies ($\hbar\omega \approx 0.5$ MeV), the model predicts a secondary alignment event. This high-spin feature is attributed to the alignment of a proton pair, occurring only after the rotational energy becomes sufficient to overcome the stabilizing influence of the $Z=72$ shell gap.

\begin{center}
\includegraphics[width=9.2cm]{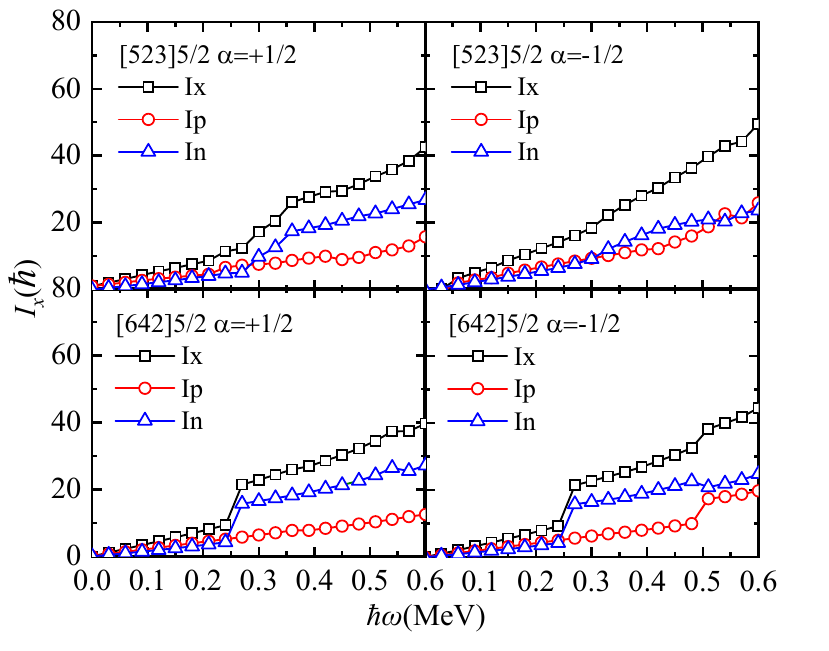}
\figcaption{Calculated proton and neutron alignment contributions for the $[523]5/2$ and $[642]5/2$ bands in $^{169}_{\;\;\;72}\text{Hf}_{97}$ extracted from the TRS minima. 
                                                                                                                           \label{Fig_IpIn}
}
\end{center}

The microscopic mechanism driving the observed alignment and subsequent signature inversion is understood in the distinct shape trajectories of the $[523]5/2$ and $[642]5/2$ bands. The shape evolution of $^{169}_{\;\;\;72}\text{Hf}_{97}$ can be extracted from the TRS minima, which are plotted against rotational frequency in Figs.\ref{Fig_b2}-\ref{Fig_b4}. 

As illustrated in Fig.~\ref{Fig_b2}, the $\alpha = -1/2$ branch of the $[523]5/2$ configuration remains relatively rigid at a high quadrupole deformation ($\beta_2 \approx 0.32$). This structural stability is attributed to the $Z=72$ proton subshell gap identified in the single-particle spectra (Fig.~\ref{Fig_SPE}), which effectively "locks" the core at this deformation.
In contrast, the $\alpha = +1/2$ signature partner exhibits a dynamic shape polarization, migrating toward a lower quadrupole deformation ($\beta_2 \approx 0.20$) while simultaneously developing a significant hexadecapole moment ($\beta_4 \approx 0.05$, Fig.~\ref{Fig_b4}). This unique hexadecapole "stretch" is restricted solely to the $[523]5/2, \alpha = +1/2$ branch. This specific $(\beta_2, \beta_4)$ geometry lowers the $i_{13/2}$ neutron intruder levels relative to the Fermi surface, facilitating a prompt, "hexadecapole-assisted" alignment. Because this alignment occurs early and rapidly for the $+1/2$ branch compared to the rigid $-1/2$ partner, it serves as the primary catalyst for the signature inversion observed at $I = 35/2$.

\begin{center}
\includegraphics[width=9.0cm]{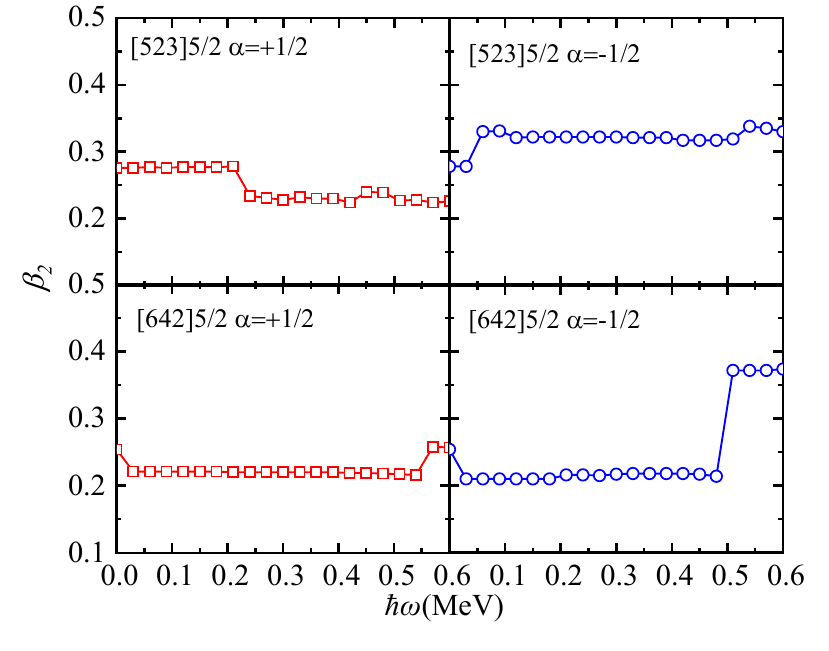}
\figcaption{Evolution of the calculated equilibrium deformation parameters quadrupole deformation $\beta_2$ as functions of rotational frequency $\hbar\omega$ for $^{169}_{\;\;\;72}\text{Hf}_{97}$. Results are compared for the $[523]5/2$ (above) and $[642]5/2$ (bottom) signature partner bands. Squares and circles represent $\alpha = +1/2$ and $\alpha = -1/2$ signatures, respectively. 
                                                                                                                           \label{Fig_b2}
}
\end{center}

\begin{center}
\includegraphics[width=9.0cm]{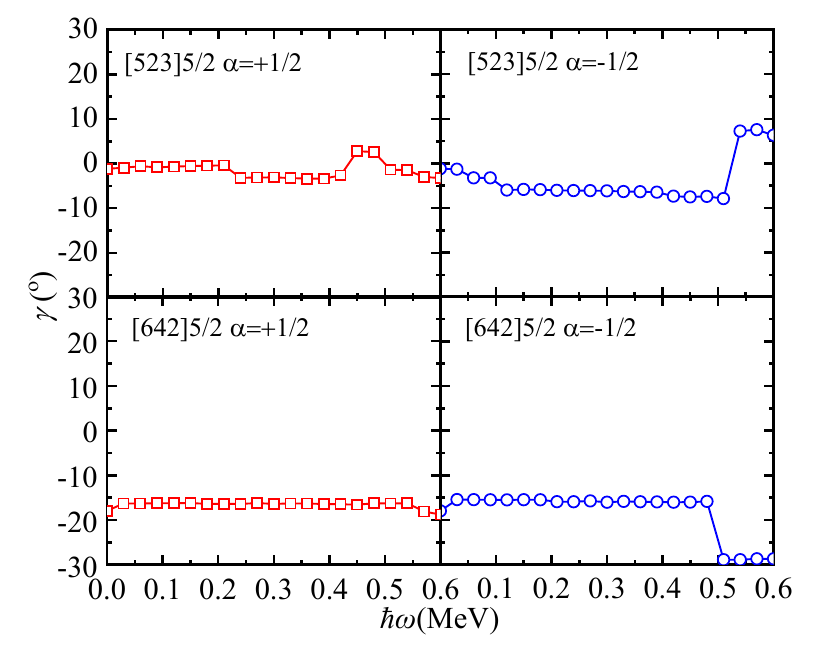}
\figcaption{Calculated equilibrium triaxiality $\gamma$ as a function of $\hbar\omega$ for $^{169}_{\;\;\;72}\text{Hf}_{97}$. For the $[523]5/2$ configuration, both signatures exhibit significant $\gamma$-softness; the $\alpha = +1/2$ branch fluctuates within a narrow range around the axial limit ($-5^\circ$ to $+5^\circ$), while the $\alpha = -1/2$ branch shows a broader transition from $-10^\circ$ to $+10^\circ$ at high frequency. In contrast, the $[642]5/2$ signatures maintain a more stable triaxial shape near $\gamma \approx -18^\circ$. 
                                                                                                                           \label{Fig_ga}
}
\end{center}

\begin{center}
\includegraphics[width=9.0cm]{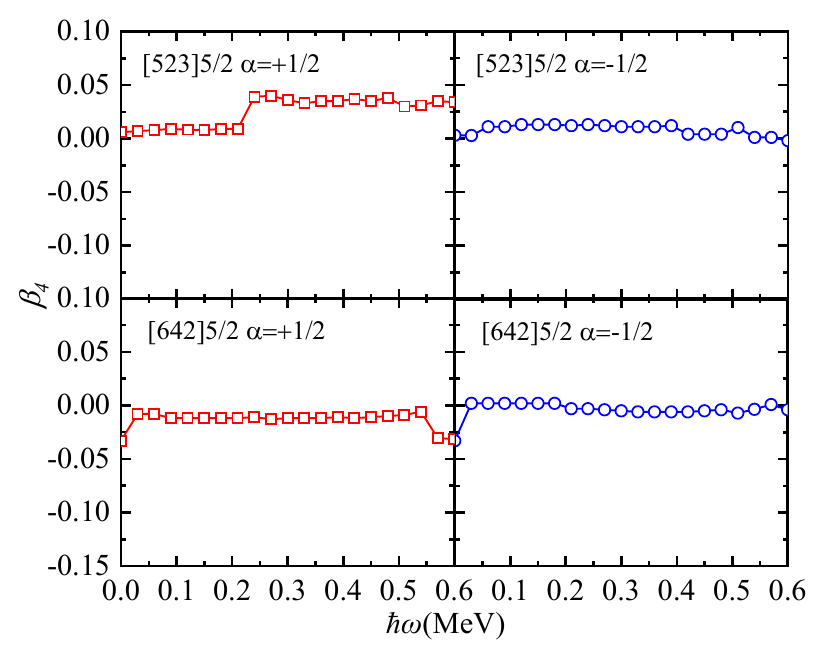}
\figcaption{Equilibrium Hexadecapole deformation $\beta_4$ versus rotational frequency $\hbar\omega$ for $^{169}_{\;\;\;72}\text{Hf}_{97}$. The calculations indicate that hexadecapole polarization is uniquely restricted to the $\alpha = +1/2$ branch of the $[523]5/2$ band, while the $[642]5/2$ band and the $[523]5/2$ $\alpha = -1/2$ branch remain hexadecapole-stable at $\beta_4 \approx 0$. 
                                                                                                                           \label{Fig_b4}
}
\end{center}

The evolution of triaxiality (Fig.~\ref{Fig_ga}) further distinguishes the rotational response. While the $[523]5/2$ ($\alpha = -1/2$) band maintains near-axial symmetry at low frequencies ($\gamma \approx 0^\circ$), it enters a $\gamma$-soft regime at high spin, with the trajectory fluctuating between $-10^\circ$ and $+10^\circ$. 
This excursion into positive $\gamma$ values is a critical factor in the inversion process, as it energetically favors the $\alpha = -1/2$ signature in the rotating frame, thereby stabilizing the inverted signature order beyond the crossing point~\cite{Stephens1975,Aryaeinejad1984}. 
The $[642]5/2$ ($i_{13/2}$) band follows a fundamentally different structural path. Throughout the observed frequency range, the band maintains a remarkably consistent shape near $\beta_2 \approx 0.20$ and a robust triaxiality of $\gamma \approx -18^\circ$. This shape stability ensures that the $i_{13/2}$ neutron alignment occurs symmetrically for both signature partners, explaining why the signature splitting remains large and no inversion occurs. Furthermore, our calculations predict a secondary alignment event at $\hbar\omega \approx 0.5$ MeV. This "shape-jump," characterized by a simultaneous increase in quadrupole deformation ($\beta_2 \approx 0.38$) and triaxiality ($\gamma \approx -30^\circ$, Fig.~\ref{Fig_ga}), signals a theoretical crossing into a highly deformed proton-aligned configuration. While currently beyond experimental limits, this predicted transition suggests a significant structural redistribution at high angular momentum once the $Z=72$ shell gap is overcome.

\end{multicols}

\section{Summary}
\begin{multicols}{2}
In this work, the high-spin rotational structures of $^{169}_{\;\;\;72}\text{Hf}_{97}$ were investigated using Total Routhian Surface (TRS) calculations based on the pairing-deformation self-consistent Cranked Shell Model (CSM). The primary objective was to elucidate the microscopic origin of the signature inversion observed in the $[523]5/2$ ($h_{11/2}$) band at $I = 35/2$. Our analysis demonstrates that this inversion is a complex dynamical process driven by the interplay between signature-dependent shape polarization and the stabilizing influence of core shell gaps.

The single-particle spectra reveal a prominent proton subshell gap at $Z=72$ for $\beta_2 \approx 0.35$. This gap acts as a "locking" mechanism for the proton core, effectively suppressing proton alignment at low-to-medium rotational frequencies and allowing neutron-driven dynamics to dominate the signature staggering. Within the $[523]5/2$ band, the calculations indicate a significant shape bifurcation: while the $\alpha = -1/2$ branch remains rigid at high deformation ($\beta_2 \approx 0.32$), the $\alpha = +1/2$ branch undergoes a transition toward a smaller quadrupole deformation ($\beta_2 \approx 0.20$) and an enhanced hexadecapole moment ($\beta_4 \approx 0.05$). This structural shift in the $\alpha = +1/2$ branch facilitates a prompt $i_{13/2}$ neutron alignment at $\hbar\omega \approx 0.3$ MeV. The resulting divergence in shape polarization between the two signature partners leads directly to a crossing of the Routhians, manifesting as the observed signature inversion.

In contrast, the $[642]5/2$ ($i_{13/2}$) band maintains a consistent, triaxial shape ($ \beta_2 \approx 0.20, \gamma \approx -18^\circ$) for both signatures throughout the observed range. This shape stability ensures that the $i_{13/2}$ neutron alignment occurs symmetrically in both branches, which preserves the large signature splitting and prevents an inversion, in excellent agreement with experimental data. Furthermore, the model predicts a high-frequency proton alignment event for the $[642]5/2$ ($\alpha = -1/2$) configuration at $\hbar\omega \approx 0.5$ MeV. This event is accompanied by a sharp transition to a highly deformed state ($\beta_2 \approx 0.38, \gamma \approx -30^\circ$), providing a clear objective for future high-precision gamma-ray spectroscopy experiments at the extremes of angular momentum.

This work highlights that signature inversion in the rare-earth region is not merely a single-particle artifact but is deeply coupled to the evolution of the nuclear shape and the specific occupation of deformed shell gaps.

\end{multicols}

\vspace{-1mm}
\centerline{\rule{80mm}{0.3pt}}
\vspace{2mm}

\begin{multicols}{2}

\end{multicols}
\clearpage

\end{document}